\documentclass[3p,times,twocolumn]{elsarticle}
\biboptions{comma,sort&compress}
 
\usepackage{graphicx}
\usepackage{amsmath}
\usepackage{here}
\usepackage{ecrc}
\volume{00}

\firstpage{1}

\journalname{Nuclear and Particle Physics Proceedings}
\runauth{Fernando S. Navarra}

\jid{nppp}
\jnltitlelogo{Nuclear and Particle Physics Proceedings}
\usepackage{amssymb}
\usepackage[figuresright]{rotating}

\begin{document}

\begin{frontmatter}

%%
%%%%%%%%%%%%%%%%%%%%%%%%%%%%%%%%%%%%%%%%%%%%%%%%%
%\begin{document}
\title{Production of  $T^+_{cc}$  in heavy ion collisions} 
 
\cortext[cor0]{Talk presented at QCD22 (04-07/07/20222, Montpellier - FR). }

\author[label1]{L. M. Abreu}
\address[label1]{ Instituto de F\'isica, Universidade Federal da Bahia, \\
Campus Universit\'ario de Ondina, 40170-115, Bahia, Brazil}
\ead{luciano.abreu@ufba.br}

\author[label2]{F. S. Navarra}
\address[label2]{Instituto de F\'{\i}sica, Universidade de S\~{a}o Paulo,  
Rua do Mat\~ao, 1371 \\ CEP 05508-090,  S\~{a}o Paulo, SP, Brazil}
%\cortext[cor1]{}
\ead{navarra@if.usp.br}

\author[label2]{M. Nielsen}
%\address[label3]{Instituto de F\'{\i}sica, Universidade de S\~{a}o Paulo,
%Rua do Mat\~ao, 1371 \\ CEP 05508-090,  S\~{a}o Paulo, SP, Brazil}
\ead{mnielsen@if.usp.br}

\author[label1]{H. P. L. Vieira}  
%\address[label4]{ Instituto de F\'isica, Universidade Federal da Bahia, \\
%Campus Universit\'ario de Ondina, 40170-115, Bahia, Brazil} 
\ead{hildeson.paulo@ufba.br}

\pagestyle{myheadings}
\markright{ }
\begin{abstract}
\noindent
We study the production of heavy multiquark states in the heavy ion collisions
performed at the LHC. We assume that they are produced at the end of the 
quark-gluon plasma phase and then interact with light hadrons during the hadron
gas phase. We use the coalescence model to compute the initial multiplicities
and effective Lagrangians to describe the interactions. We find that the initial 
multiplicity of molecules is two orders of magnitude larger than that of
tetraquarks. The interactions in the hadron gas reduce the number of molecules
and increase the number of tetraquarks by a factor two in each case. At kinetic
freeze-out, the difference is still very large. 
%% keywords
%\ccode{Pac numbers: 11.55.Hx, 12.38.Lg, 13.20-Gd, 14.65.Dw, 14.65.Fy, 14.70.Dj}  
\end{abstract}
\end{frontmatter}
%%%%%%%%%%%%%%%%%%%%%%%%%%%%%%%%%%
%\end{document}
%%%%%%%%%%%%%%%%%%%%%%%%%%%%%%%%%%
%\vspace*{-1.5cm}
\section{Introduction}
%\vspace*{-0.25cm}
 %\nin
%%%%%%%%%%%%%%%%%%%%%%%%%%%%%%%%%%%
The number of exotic hadrons keeps continuously growing. The most
recently observed exotic states were announced during this conference
\cite{4338} \footnote{The $J/\psi - \Lambda$ bound state was observed in 
the $B \to J/\psi \Lambda \bar{p}$ decay, as first proposed in \cite{brona}. }
as well as the most recent review article on the subject \cite{mapi22}. A more
complete review can be found in \cite{epp17}.  
With this new experimental knowledge we can expect to
deepen our understanding of both the QCD and hadronic interactions. Since the
observation of the first exotic hadron, the $X(3872)$, in 2003, one of the
main questions in the field is: which is the internal structure \cite{nnn}
and spatial configuration of these multiquark states ? The two most accepted
configurations are loosely bound hadron molecules and compact tetraquarks.  
This question gained a new interest after the recent discovery of the 
$T_{cc}^+$ \cite{nnl07}.  

We may try to answer this question using the heavy ion collisions collisions
performed at LHC, as suggested in \cite{exhic}. In these collisions,
the number of produced charm quarks is much larger than in any other type of
collision. Moreover, even if the environment is much less clean than in
$e^+ \, e^-$ or in $p \, p$ collisions,  modern techniques allow for the
reconstruction of these states.  The $X(3872)$ has been recently observed by
the CMS collaboration in $Pb-Pb$ collisions \cite{cms}. This opens a new chapter
in heavy hadron spectroscopy. 

The striking new feature of heavy ion collisions is the formation of the quark
gluon plasma (QGP). It introduces a new way of formation of hadrons: coalescence
during the back transition to the hadron phase. This is, of course, a
non-perturbative process, which can only be calculated with the help of models,
such as the one presented in \cite{exhic}. This model yields an interesting
prediction: molecules are much more abundantly produced than tetraquarks! This
is due to the fact that, at the hadronization time, the quark system has a
density  which is comparable to normal hadronic matter, in which three quarks
occupy the typical volume of a baryon and a quark-antiquark pair occupies the
volume of a meson. At this point, two formed mesons have a typical separation of
a few fermi and can easily form a molecule. In sharp contrast, the typical
separation between the quarks is much larger than the size of compact
tetraquarks ($ \simeq 0.2 - 0.5$ fm). The quantitave implementation of these
ideas requires several assumptions concerning the quark distribution in the
QGP, the interactions in the bound state and its wave function. The numerical
results confirm the intuitive expectation and the conclusion is that molecules
would be about hundred times more abundant than tetraquarks. This prediction
was first published many years ago in (see references in \cite{exhic}) but a
proper treatment of the hadronic interactions of the multiquark states was only
developed in the most recent years. We are going to briefly discuss
the subject in the next sections. 

\section{Interactions in a hadron gas} 

After being formed during the hadronization, the particles have to live for
some time ($\simeq 10$ fm) in a hot hadron gas, where they interact with light
mesons. These interactions can be described by effective Lagrangians. Using this
approach, we have studied the interactions of the $X(3872)$ \cite{nosx},
the $Z_b$ \cite{zb}, the $J/\psi$ \cite{psi}, the $\Upsilon$ \cite{ups} and,
more recently, the $T_{cc}$ \cite{annv22}. Even though the Lagrangian may be the
same, molecules and tetraquarks interact in a different way. The difference
comes from their different sizes and it is encoded in the form factors and
coupling constants. For tetraquarks, these quantities can be calculated with
QCD Sum Rules (QCDSR) \cite{svz,nari1}. For molecules there are models.

In \cite{annv22} we compared the interactions of a ``pure'' molecular state with 
the interactions of a ``pure'' tetraquark state in a hadronic medium.

The
molecular model was taken from \cite{lee18}. In this ``quasi-free'' model of the
$T_{cc}^+$, the $D$ and the ${D}^*$ are considered to just ``fly'' together.
Both the binding energy and the interaction between these constituent mesons are 
neglected. They are destroyed by any elastic scattering between a $\pi$ and one
of the constituent charm mesons. The advantage of this approach is that the only
Lagrangian required is the one of the $D^* D \pi$ vertex. Fortunately the
$D^* \to D \pi$ decay is measured and the coupling constant is known. As in all
effective Lagrangian approaches, we need a form factor which contains a free
parameter (the cut-off $\Lambda$). This is the main source of uncertainty.

In \cite{annv22} we also developed the tetraquark approach. In this case the
$T_{cc}^+$ is a degree of freedom, a new field appearing in the Lagrangians.
We have several interaction vertices, described by Lagrangians. All of them
were already known from previous studies, where with the help of QCDSR, we had
computed the form factors and coupling constants \cite{bcnn12}. The new
ingredient is the $T_{cc}^+ \, D \, D^*$ vertex. In \cite{annv22} we performed
the QCDSR study of the relevant three-point function and determined the form
factor and the coupling constant in this vertex. 

\begin{figure}[!ht]
\includegraphics[{width=7.0cm}]{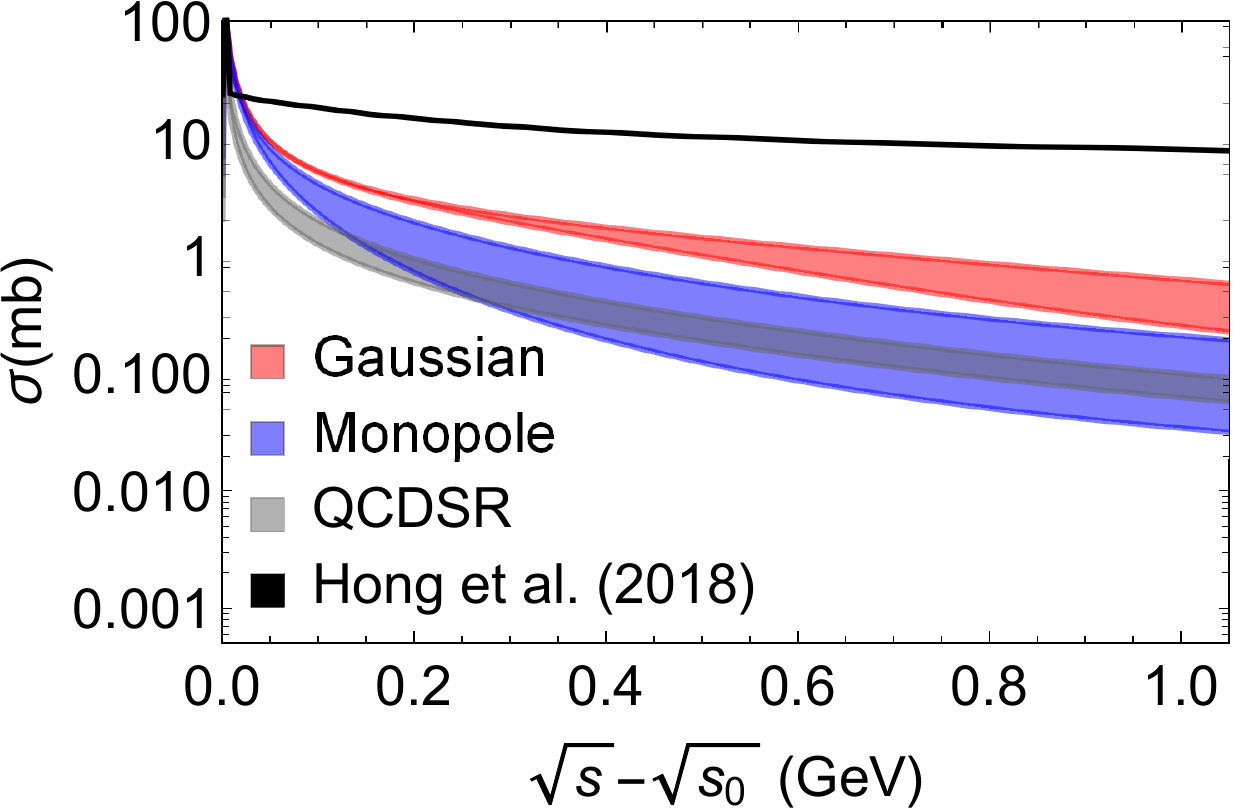} %crosscomp.eps}
\caption{Cross section of the process $T_{cc}^+ \, \pi \to X$.
The ``quasi-free''  molecular is represented by the black solid line
(the cut-off was chosen to be $\Lambda = 1$ GeV. The other bands show the
results obtained with the tetraquark model. The (wider) blue and red bands show
results obtained with empirical gaussian and monopole form factors (the
boundaries of the bands are defined by the cut-off choices going fom $1$ to
$2$ GeV. The central
(thinner) grey band shows the results obtained with form factors derived from
QCDSR. }
\label{crosscomp}
\end{figure}

With these results we could compute the absorption cross sections of
$T_{cc}^+$ by pions in both the molecular and tetraquark approach. The
comparison is shown in
Fig.~\ref{crosscomp}. As we might expect, the molecular cross section is much
larger. It is then much easier (by almost two orders of magnitude) to destroy
a molecule than a tetraquark. In hadronic interactions it is also easy to
create the $T_{cc}^+$ in the inverse processes, such as, for example, 
$ D^* \, D \to T_{cc}^+ \, \pi$. However, another robust conclusion
of \cite{annv22}
valid both for molecules  and tetraquarks, is that the cross section for
$T_{cc}^+$ production is ten times smaller than the cross section for
$T_{cc}^+$ absorption.
%\vfill\eject
%%%%%%%%%%%%%%%                                             

Apart from the numbers, it is interesting to observe how the QCDSR calculation
reduces the uncertainties. Moreover, the uncertainties in QCDSR come from
quantities, such as  quark masses and condensates, which can be constrained
by  other experiments.

\section{Hadron multiplicities and system size dependence} 
   
\begin{figure}[!ht]
\includegraphics[{width=7.0cm}]{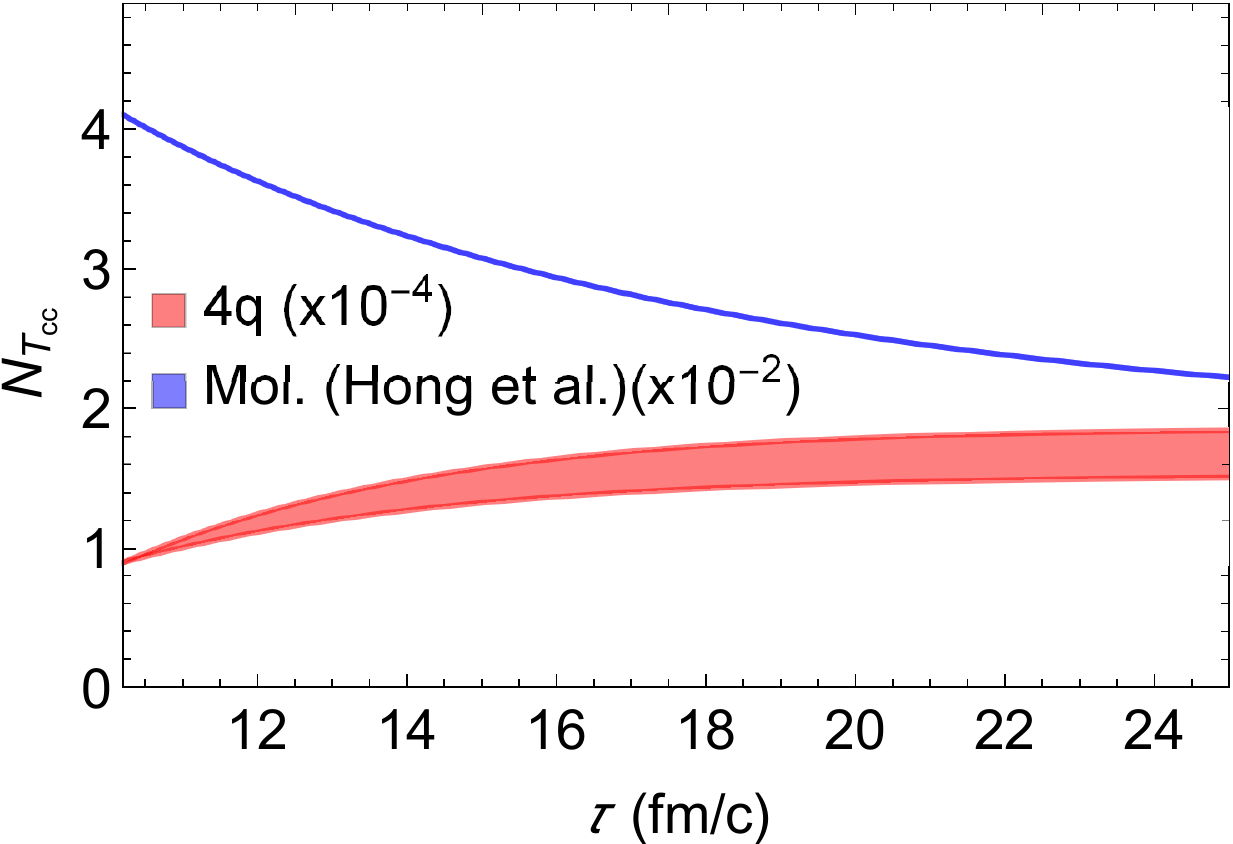} %Tccf-fig5a.eps}
\includegraphics[{width=7.0cm}]{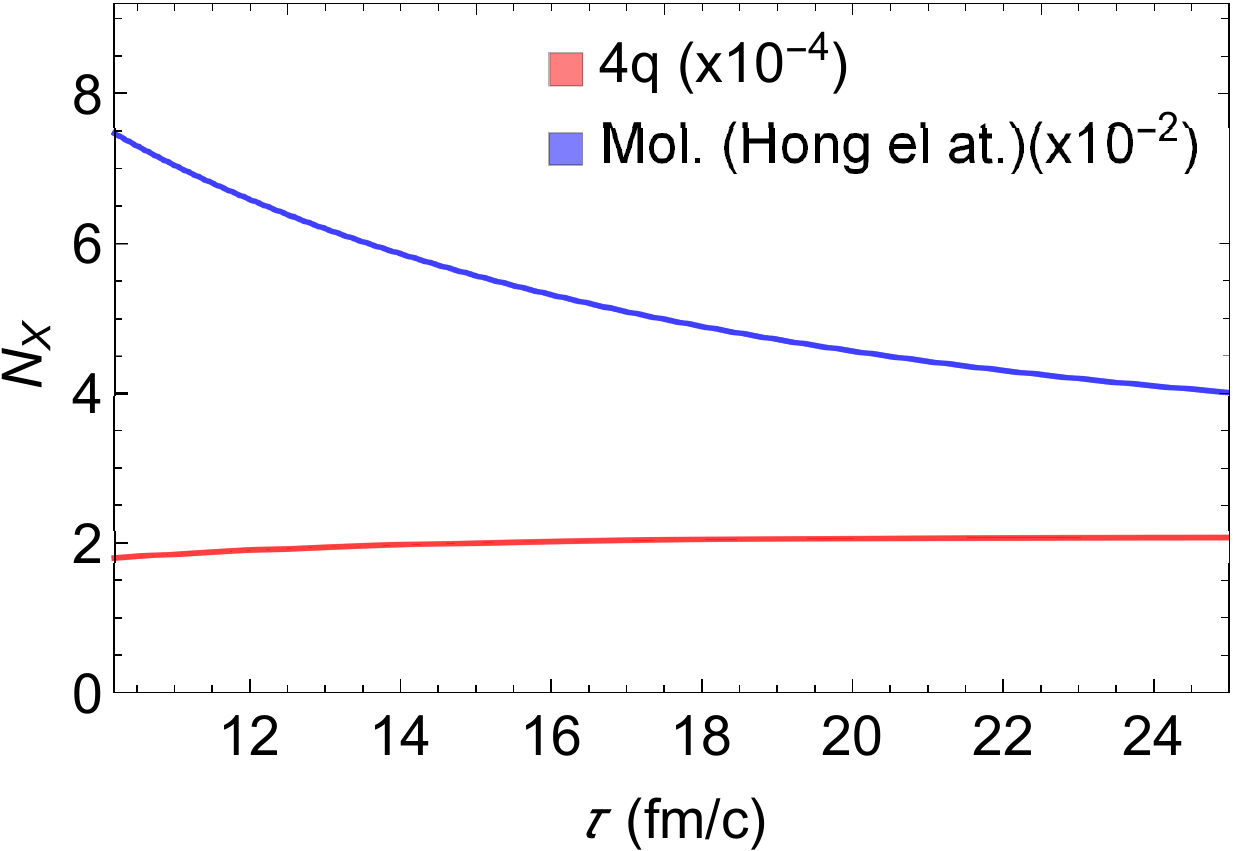} %Tccf-fig5b.eps}
\caption{Time evolution of the $T_{cc}$ (upper panel) and
  $X(3872)$ (lower panel) abundances as a
  function of the proper time in central $Pb-Pb$ collisions at
  $\sqrt{s_{NN}} = 5.02$ TeV. Details can be found in \cite{anv22}. }
\label{timevol}
\end{figure}

Having computed the cross sections, we can compute the corresponding thermal
cross sections and write rate equations for the abundances $N_{T_{cc}}$
and $N_{X(3872)}$. This was done in \cite{anv22} and the result is shown in
Fig.~\ref{timevol}. The initial conditions are given by the coalescence model
\cite{exhic}. Comparing the two panels, we see that, not surprisingly, the
multiplicities of $T_{cc}$ and $X(3872)$ evolve with time in a quite similar
way and the differences come mostly from the different initial conditions. It
is also natural the fact that the multiplicities decrease for molecules and
increase for tetraquarks, since the latter are much less abundant. Finally,
looking at these curves we can observe that the particle abundances are
approaching  constant values. This would mean that they are reaching the
chemical equilibrium. Interestingly, the approach to equilibrium depends on the
configuration and tetraquarks seem to be already in equilibrium at the end of
hadron gas phase in the considered heavy ion reaction, in contrast to the
molecular configuration.  This raises the possibility of discriminating between
the two configurations. In the example shown in the figure, if the observed value
of the abundance would be close to the value predicted by the Statistical
Hadronization Model (SHM)\cite{shm}, then we could conclude that the particle
is a tetraquark. 
%\vfill\eject
%%%%%%%%%%%%%%%

\begin{figure}[!ht]
\includegraphics[{width=7.0cm}]{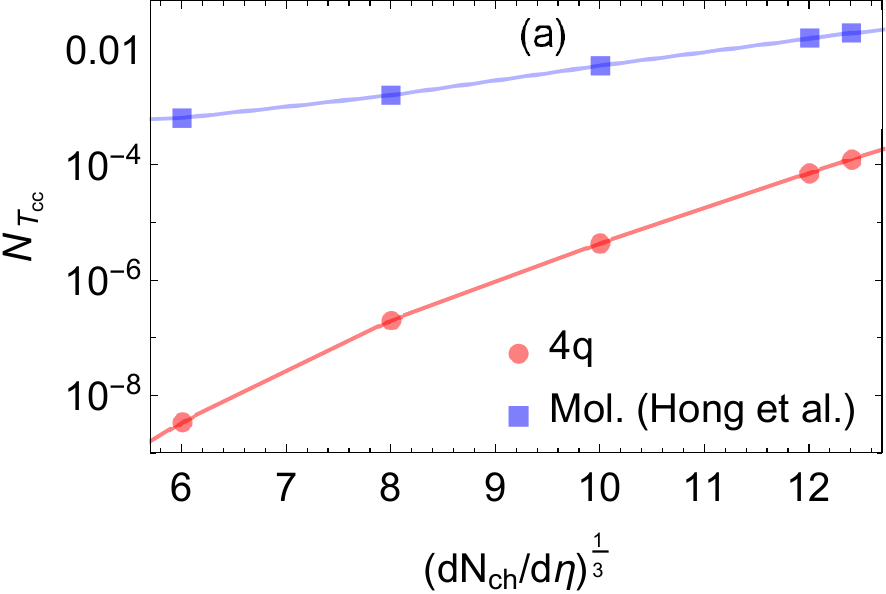} %Tcc-dndeta.eps}
\vspace*{0.5cm}  
\includegraphics[{width=7.0cm}]{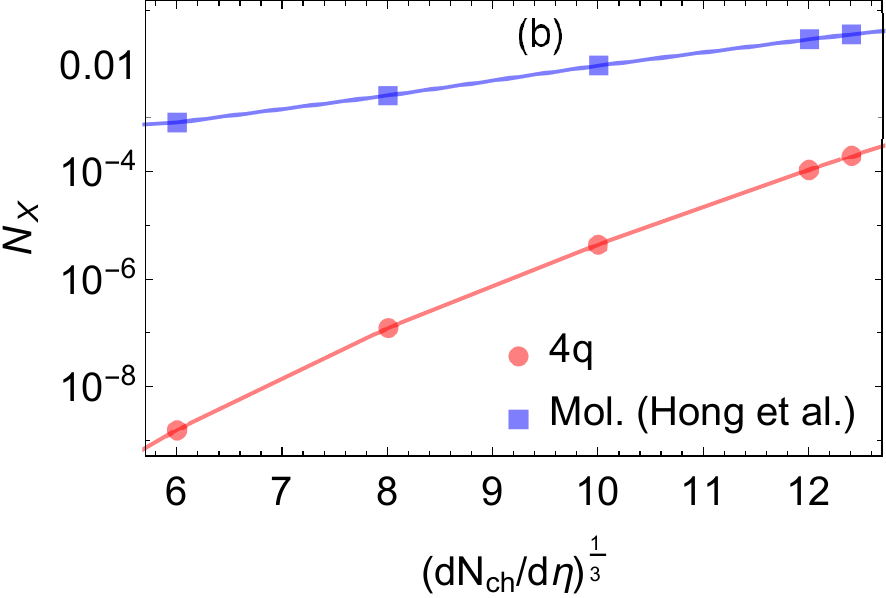} %X-dndeta.eps}
\caption{Multiplicity as a function of
  $\mathcal{N}=(d\, N / \, d\eta)^{1/3}$. Details can be found
  in \cite{anv22}.} 
\label{N-dndeta} 
\end{figure}

In order to make predictions for the measurable particle abundances we
would have to combine our microscopic description of the interactions with
a full simulation of the underlying event, i.e., we would need a detailed
code able to generate the whole heavy ion collision. However, using various
pieces of experimental information, it is possible to establish a connection
between the abundance of particles given by the coalescence model and the
system size, represented by the charged particle density at mid-rapidity
$d N /d \eta \,  (\eta =0)$.  This was done in \cite{anv22} and the result is
shown in Fig.~\ref{N-dndeta}. As it can be seen, as we move to smaller systems
(smaller values of $\mathcal{N}$) the difference of molecule and tetraquark
yields increases. This happens because smaller systems live for shorter times
and there is no time for the absorption-creation dynamics to be effective enough
and reduce the difference in the yields. In this sense Fig.~\ref{N-dndeta}
is a consequence of Fig.~\ref{timevol}. The pattern of separating curves for
smaller values of $\tau$ translates into the same pattern for smaller values
of $\mathcal{N}$. The predictions of Fig.~\ref{N-dndeta} can be compared with
future data. Of course, we have to keep in mind that, as the system becomes
smaller, the whole description based on QGP formation looses validity. 

\section{Discussion and conclusion}

The picture developed above received some criticism, which we review
below. 

A  first  criticism was addressed to the coalescence model
\footnote{We deeply grateful to S.H. Lee and to R. Rapp, for
  formulating this criticism so clearly.}.
It assumes that if the overlap
between the quark spatial distribution in the final stage of the QGP and the
size of the formed hadron is large, then the abundance of this hadron will be
large. If this was true, the yield of $\psi(2S)$ ($r \simeq 0.8$ fm) would be
larger than the yield of $J/\psi$ ($r \simeq 0.4$ fm). Data show the opposite!
Although this argument is correct, it would be necessary to
develop a theory for the interaction of the $\psi(2S)$ with light mesons. It
might well be the case that these interactions would be responsible for the
suppression of this particle. Moreover the sizes are not so different as in the
case of multiquark states and the
initial multiplicities given by the coalescence model would certainly favor
the $\psi(2S)$  over the $J/\psi$, but not by a factor 100 !

A second criticism is the following. 
In heavy ion collisions we also observe deuterons, which are the prototype
of a very weakly bound hadronic system. Also, the predictions of the SHM for 
particle yields have been, to a large extent, confirmed.
Even for deuterons, as shown in \cite{shm}.
This suggests that the hot system formed in heavy ion collisions reaches
chemical equilibrium. When this happens the production yield is controled
solely by the mass, chemical potential and temperature. The details of the
initial conditions and interactions are ``forgotten''. How to reconcile the
microscopic transport dynamics with the success of the SHM, which knows nothing
about this dynamics ?

If the multiquark system is a molecule, one may assume that, if it is
formed in quark-hadron transition, it is completely destroyed, absorbed by
the medium. Molecule production (such as the deuteron) happens only at the
last moment of the hadron gas phase by hadron coalescence. Then, if the chemical
and kinetical freeze-out temperatures are close enough, the predictions of the
SHM and the coalescence model may coincide. However, this has to be checked.

If the multiquark system is a tetraquark, it may be formed already inside the
QGP and inherit its thermal properties \cite{rapp}.
Then it would traverse the hadron gas
almost without interacting (in a sort of ``color transparency'') and emerge at
the end of the collision remembering its temperature. Again, this
almost non-interacting behavior of tetraquarks has to be proven. Our results, 
shown in Fig.~\ref{crosscomp}, suggest that the interaction cross section is
sizeable. 

%In this discussion it also important to remember that the observed $X(3872)$
%have a too large transverse momentum  ($10 < p_T < 25 $ GeV) to come from a  
%hadronic QGP - Hadron Gas environment.  This kind of $p_T$ may be related to
%jet dynamics. The deuterons are produced with low $p_T$ ($0 < p_T < 5 $ GeV). 
%In both cases, one can not just ``decree'' that the hadronic interactions  
%behave as we described in the paragraph above. It would be necessary to
%confirm this statement by doing the kind of calculation that we do.

To conclude, we believe that heavy ion collisions mark the beginning of a new
era for the study of exotic hadrons. From the theoretical side,
it is necessary to have the best possible description of the interactions of
the exotic mesons in the hadron gas.

%\vfill\eject                                                                   
%%%%%%%%%%%%%%%
\vspace*{0.5cm}

\end{document}